\documentclass{hep99}
\begin{document}

\title{Fine tunings and quark masses: Phenomenology of multiple
domain theories}

\author{John F. Donoghue }
%
% Use of footnote symbols, footnoted material after \maketitle 
%\author{A J Cox$^1$\dag\ and Jim Revill$^2$\ddag}

\address{ Department of Physics and Astronomy, University of Massachusetts,
Amherst MA 01003 \\ and \\ TH Division, CERN, Geneva \\[3pt]
E-mail: {\tt donoghue@phast.umass.edu}}

\abstract{This talk describes some of the consequences for particle 
phenomenology of the hypothesis that the physical parameters may
vary in different domains of the universe.} 

\maketitle

\section{Introduction}

This talk\footnote{Talk presented at the 1999 European Physical Society
HEP Conference, Tampere, Finland, July 1999}
 is a mini-review of a possible pathway in the search for the 
fundamental theory. This approach is quite distinct from the usual
directions taken in searching for new theories, and hence may appear
a bit odd at first. However, it may also lead to new possibilities 
and could prove useful.

The basic hypothesis is that there exist different domains in the 
universe where (at least some of)
the parameters of the underlying theory can take on
different values. We would live entirely within one such domain
and, under the assumption of inflation, we would not see any variation
within this domain nor would we have access to other domains. This
multiplicity of parameters and domains 
is in strong contrast with the usual assumption that if we work hard 
enough, we can uncover the theory whose unique ground state determines
our world.

This may not be as crazy as it sounds at first. An effect like this
can occur in chaotic inflation\cite{Linde}, 
where scalar fields can get frozen at
random values if their potential is flat enough. It is also
a 
conceivable outcome in string theory where there are continuous 
families of ground state solutions, and we have little insight 
as to how one ground state is selected or preserved. However, it is
enough to have occurred in one physical theory, such as chaotic inflation,
to need to take the general idea seriously as a possibility.

The idea is also not as empty as it first sounds. Clearly
it tell us that the some specific parameters that we see may 
not be uniquely 
predictable. However, as described below, there is still some 
information contained in those parameters. Moreover, the hypothesis
can suggest that certain problems, such as fine tuning problems, are less
serious than they first appear and thus motivate new approaches to the 
exploration of fundamental theories.

\section{Weinberg and the cosmological constant}

Weinberg has made a physical calculation that is relevant for 
this hypothesis\cite{Weinberg, Martel}. 
He notes that for most values of the cosmological
constant the universe is extreme and sterile, either living an
extremely short time of order the Planck scale or expanding too fast
for matter to ever clump. He calculates the range of the cosmological
constant that allows galaxies to clump, and finds that it 
very small. This then leads to a natural constraint on
our domain - out of all possible domains we would only find ourselves 
in a domain such that matter clumps. In turn, this leads to a 
consistency check on whether it is reasonable to think that this constraint
is the main explanation for the smallness of the cosmological
constant or whether other explanations must be sought. If the observed
value of the constant is very much smaller than the allowed range, we
would expect that another mechanism is needed to make it so small. 
However if the value is typical of the range then no extra explanations
are needed within the class of multiple domain theories. 

The actual range and the mean value have been estimated\cite{Martel}, and 
the interesting feature is that the newly observed value of the 
cosmological constant is reasonably typical of the viable range.
A zero value of the cosmological constant is already extremely difficult
to understand theoretically. A non-zero value of this extremely tiny 
magnitude is even harder to understand by a dynamical mechanism. 
If the observed cosmological constant is correct, it finds a natural home
in multiple domain theories and, by itself, is a reason to 
take this hypothesis seriously.

There are two recent developments related to Weinberg's result.
Tegmark and Rees\cite{Tegmark} have 
pointed out that the initial strength of density
perturbations, Q, also enters into the calculation of gravitational 
clumping, They show there is a limited viable region in the two-dimensional
space of Q and $\Lambda$, thus generalizing 
Weinberg's constraint. In addition, Garriga and Vilenkin\cite{Garriga} 
has pointed out that
Weinberg's assumption of a flat weight for the distribution
in the cosmological constant may not hold in various Higgs models, and
that this weight can lower the mean viable value. 
Both of these represent interesting developments
of Weinberg's original calculation, and do not diminish the attractiveness
of the general idea. 

\section{Fine tuning of the the Higgs mass parameter}

The other great fine-tuning problem that motivates particle 
physicists is that of the Higgs vacuum expectation value.
A similar constraint can be calculated in this case. Here
the assumption is that the existence of complex elements is
a natural constraint for the domain that we find ourselves in. 
That is, domains in which there is only one element do not have
the complexity needed for life of any sort. My collaborators
and I\cite{Higgs} 
have tried to estimate this viable range for the Higgs mass
parameter, under the assumption that all of the other dimensionless
parameters of the Standard Model do not change.

The basic physics is that the Higgs vev controls quark masses and,
if the quark masses increase a modest amount, complex matter ceases to
be exist in the universe. The first problem is the unbinding of deuterium
as the pion gets slightly heavier. Deuterium is needed in all of the 
mechanisms for element production. However, a more serious constraint 
occurs at a vev about five times that observed, when the neutron 
becomes heavier than the proton by enough that all nuclei are 
unstable to decay to free protons. This leaves a universe of protons only.
At much larger values of the vev, the $\Delta^{++}$ becomes the only element
but there is still not enough complexity for life. Thus out of
the whole range for the Higgs vev, the observed value is
reasonably typical of the viable range.

This was done under the assumption that the other constants 
have been held fixed. However, it is likely to be a reasonably 
robust conclusion. The most general way to state the result is that
the existence of complex elements requires the weak scale and the
QCD scale to overlap. The quark masses are 
manifestations of the weak scale. In the real world, some of these 
masses are below the QCD scale and some are above. Complex elements
only arise through the interplay of the QCD scale and the quark 
masses, which allows more than one hadron to have masses close
enough to each other to provide variation in the nuclei. 
(Electromagnetic effects at order $\alpha\Lambda_{QCD}$ also are
important in determining the pattern of nuclei.) The overlapping of the
QCD scale and the weak scale is a puzzle for fundamental theories
which is distinct from the issue of fine-tuning. 
In the context of low-energy supersymmetry, if it exists, these considerations
can be rephrased as the answer to the question of why, out of all the 
available parameter space, SUSY breaking takes place so close to the QCD 
scale.

There has recently been a work which helps to strengthen this 
result by pointing out that the production of the carbon 
would not have been possible if the Higgs vev was
modestly smaller than observed.\cite{Sher}

\section{Comments on anthropic constraints}

The above constraints are examples of reasoning that goes under the
name of ``the anthropic principle''. 
There is a large and varied literature on anthropic ideas. This 
includes works of a technical nature, of which an excellent
survey is found in the book of 
Barrow and Tipler\cite{anth1}, as well as those that 
provide thoughtful
discussions\cite{anth2}. 
The treatments above provide a different emphasis 
on ideas
that appear throughout this literature, with a focus on the present 
key problems of particle physics.  
Much of the literature on anthropic ideas uses a
narrow definition of life, one centered closely on life as we
know it.  The 
analyses which I described 
attempt to choose a much looser definition of the conditions
relevant for the possibility of life (clumping of matter and the presence 
of complex elements). They also consider a much wider variation 
of the parameters, and attempt to calculate {\it typical}
values of the parameters. 

One of the criticisms of anthropic arguments is that they are 
just a way to get around making real testable predictions. 
Such abuse is always possible, but that is not really the point
of such studies. Rather, one is interested in understanding
which questions are fruitful to consider. 

Much of the research in particle theory beyond the Standard Model
is driven by the fine-tuning problems. The assumption that 
supersymmetry is present down to low energies seems to have 
permeated the field. However, this could turn out to be wrong - 
which is why we must do the experiments to test it. The present
indications of the existence of a cosmological constant should give us 
all some concern about fine-tuning arguments. Here is a de-facto
fine tuning which does not appear to be solved by having new
physics at the relevant scale. The anthropic 
considerations discussed above might be interpreted as
the possibility that the fine-tuning problems are {\em not} the most
important ones facing us\footnote{Note however that anthropic constraints cannot
``solve'' the strong CP problem. The $\Theta$ parameter is many orders
of magnitude smaller than its viable mean value, 
and we need to seek a dynamical
explanation for this.}.

\section{The weight for quark masses}

If the quark masses are also parameters that can vary in different 
domains, then attempts to predict the specific values of the masses 
will not be fruitful. However the masses that we see are not really
random. For example, there are more light masses than really heavy ones.
It is not necessarily the case that the mass spectrum should be flat
if they are variable. They may be distributed with respect to some
weight. The interesting feature is that the residual information
about the underlying theory is not in the specific masses, but in the 
weight. In such theories, the weight can be used as a test of the theory.
 
The observed weight in our domain has an intrinsic uncertainty since
we only have information on 6 quark masses and 3 lepton masses. (I am
assuming here that the physics of neutrino mass must be treated separately.)
Nevertheless, when one tries to extract the weight from the data,
it is remarkable that the uncertainty is not so great\cite{Donoghue}. 
The answer can
be summarized by saying that the weight is approximately the scale
invariant form proportional to 1/m. (More precisely, the inverse power
can vary between roughly 0.85 and 1.) If multiple domain theories are
what occurs in nature, this can be a hint as to the structure of the
correct theory. 

\section{Comments}
   
   At present, the ideas described above amount to little more than a 
``story'' about how the theory could work. There has been little effort 
devoted to dynamical mechanisms. The example
of chaotic inflation shows that it is indeed possible for physical parameters
to be fixed at a continuous range of values. However it is not known
how widespread this mechanism is in other theories. Certainly cosmology
is the primary setting to explore the effect. In cosmology, causally
disconnected 
regions in the early universe will have different conditions, and hence the
initial conditions may possibly lead to different parameters. The implementation
of these ideas in fundamental theories is an interesting challenge.

\end{document}